\documentstyle[12pt,prc,aps,psfig]{revtex}
\tighten
\begin{document}
\draft

\title{Nucleon-nucleon cross sections in neutron-rich  \\
 matter}
\author{F. Sammarruca and P. Krastev}
\address{Physics Department, University of Idaho, Moscow, ID 83844, U.S.A}
\date{\today}
\maketitle
\begin{abstract}
We calculate nucleon-nucleon cross sections in the nuclear medium with
unequal densities of protons and neutrons. We use the Dirac-Brueckner-Hartree-Fock
approach together with 
realistic nucleon-nucleon potentials. 
We examine the effect of
neutron/proton asymmetry and find that, although generally mild, it can
be significant in specific regions of the phase space under consideration. 
\end{abstract}
\narrowtext

\section{Introduction} 

 A topic of contemporary interest in nuclear physics is the investigation
 of the effective nucleon-nucleon (NN) interaction in a dense hadronic 
 environment. Such environment can be produced in the laboratory via energetic
 heavy ion (HI) collisions or can be found in astrophysical systems, particularly the   
 interior 
 of neutron stars. In all cases predictions rely heavily on the nuclear equation
 of state (EOS), which is one of the main ingredients for transport simulations
 of HI collisions as well as the calculation of neutron star properties. 

 Transport equations, such as the Boltzmann-Uehling-Uhlenbeck (BUU) equation, 
 describe the evolution of a non-equilibrium gas of 
 strongly interacting hadrons. In BUU-type models,   
 particles drift in the presence of the mean field while undergoing    
 two-body collisions, which require the knowledge of in-medium two-body cross sections.
 In a microscopic approach, 
 both the mean field and the binary collisions are calculated 
  self-consistently starting from the bare two-nucleon force.

 We present microscopic predictions of NN cross sections in symmetric and 
 asymmetric nuclear matter. In asymmetric matter, the cross section becomes
 isospin dependent beyond the usual and well-known differences between the $np$ and the $pp/nn$ 
 cases. It depends upon the total density 
 and the degree of proton/neutron asymmetry, which of course also implies that the 
 $pp$ and the $nn$ cases will in general be different from each other. In this paper,
 we will only be concerned with  
 the the strong interaction contribution to the cross section
 (Coulomb contributions to the
 $pp$ cross section are not considered). 

Asymmetry considerations are 
of particular interest at this time. The planned Rare Isotope Accelerator 
will offer the opportunity to study collisions of neutron-rich nuclei 
which are capable of 
producing extended regions
of space/time where both the total nucleon density and the neutron/proton 
asymmetry are large.                                                               
Isospin-dependent BUU transport models \cite{BUU} 
include isospin-sensitive collision dynamics through the elementary $pp$, $nn$, and $np$
cross sections and the mean field (which is now different for protons and        
neutrons).
 The latter is a crucial isospin-dependent mechanism
and is the focal point of an earlier paper \cite{SBK}. 

 In a simpler approach, the assumption is made that the transition matrix
 in the medium is approximately the same as the one in vacuum and that         
 medium effects on the cross section 
 come in only through the use of  nucleon effective masses in the phase space factors 
 \cite{PP,Gale}. Concerning microscopic calculations,  
  earlier predictions can be found, for instance, in                          
 Refs.~\cite{LM,Fuchs}, but asymmetry considerations are 
 not included in those predictions. 
 On the other hand, it is important to 
 investigate to which extent the in-medium
 cross sections are sensitive to changes in proton/neutron ratio, the main purpose of this 
 paper.
In-medium cross sections are necessary 
to study the 
mean free path of nucleons in nuclear matter and thus nuclear 
transparency. The latter is obviously related to the total reaction 
cross section of a nucleus, which, in turn, can be used to extract
 nuclear r.m.s. radii within Glauber-type models \cite{Glauber}. Therefore,  accurate in-medium 
 {\it isospin-dependent} NN cross sections can ultimately be very valuable to obtain 
 information about the size of exotic, neutron-rich nuclei.

In the next Section, after providing some details on the calculation, we present
and discuss our results. Our conclusions are summarized in Section III.

\section{In-medium $NN$ cross sections} 
\subsection{General aspects} 
Details of our application  of the Dirac-Brueckner-Hartree-Fock framework 
to asymmetric matter
can be found in Ref.\cite{AS02}. 
We choose                                  
 the Bonn-B potential \cite{Mac89} as our model for the free-space 
 two-nucleon (2N) force. 

  The nuclear matter calculation of Ref.~\cite{AS02} provides, 
 together with the EOS, 
 the single-proton/neutron potentials as well as their parametrization in 
 terms of effective masses \cite{SBK}. Those effective masses, together with 
 the appropriate Pauli operator (depending on the type of nucleons involved), 
  are then used in a separate calculation of the in-medium       
  reaction matrix (or $G$-matrix) under the desired kinematical conditions.

\begin{figure}
\begin{center}
\vspace*{0.3cm}
\hspace*{-0.5cm}
\psfig{figure=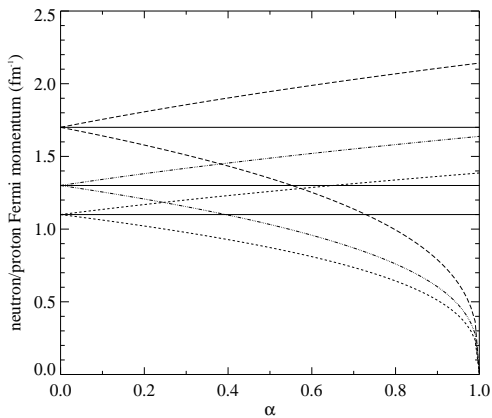,height=6.0cm}
\vspace*{0.3cm}
\caption{Increasing(decreasing) of the neutron(proton) Fermi momentum as a function of
the asymmetry parameter. The average Fermi momenta corresponding to the three
groups of curves are 1.1, 1.3, and 
1.7 fm$^{-1}$, respectively. 
} 
\label{one}
\end{center}
\end{figure}
 Our calculation is controlled by the total density, $\rho$, and                    
 the degree of asymmetry,
$\alpha=(\rho_n - \rho _p)/(\rho_n + \rho_p)$, with $\rho_n$ and $\rho_p$ the 
neutron and proton densities.                                                  
For the case of
identical nucleons, the $G$-matrix is calculated using the appropriate
effective mass, $m_i$, and the appropriate Pauli operator, $Q_{ii}$,                      
depending on $k_F^i$,            
where $i=p$ or $ n$. 
For non-identical nucleons, we use 
the ``asymmetric'' Pauli operator, $Q_{ij}$, depending on both 
$k_F^n$ and
$k_F^p$ \cite{AS02}. In Fig.~1, we show the variations of 
$k_F^n$ and
$k_F^p$ with increasing neutron 
fraction at three fixed densities, according to the relations
\begin{equation}
k_F^n = k_F(1 + \alpha)^{1/3}
\end{equation}
\begin{equation}
k_F^p = k_F(1 - \alpha)^{1/3} . 
\end{equation}
This may facilitate the 
interpretation of results later on.                                               

 In the usual free-space scattering scenario, the cross section is typically           
 represented as a function of the incident laboratory energy, 
 which is uniquely related to the nucleon momentum
 in the two-body c.m. frame, $q_0$ (also equal to the 
 relative momentum of the two nucleons), through the well-known
 formula $T_{lab} = 2q_0^2/m$.                                       
 In nuclear matter, though, 
 the Pauli operator depends also on the total 
 momentum of the two nucleons in the nuclear matter rest frame. 
 This could be defined in some average (density-dependent) manner, or kept 
 as an extra degree of freedom on which the cross section will depend.
 We will take the latter approach and                     
 examine this aspect in the next subsections. 

Another issue to consider carefully when approaching the concept of in-medium cross sections is the non-unitary nature of the interaction. 
Due to the presence of Pauli blocking,                                          
the in-medium  scattering matrix does not obey the free-space unitarity relations through
which phase parameters are usually defined and from which it is customary to          
determine the NN scattering observables.
As done in Ref.~\cite{Fuchs} but unlike Ref.~\cite{LM}, 
we calculate the cross section directly from the 
scattering 
matrix elements thus avoiding the assumption of in-vacuum unitarity and its    
consequences (such as the optical theorem). 
That is, we 
integrate the differential cross section                      
\begin{equation}
  \sigma(q_0,\rho)= \int \frac{d\sigma}{d\Omega}(q_0,\theta,\rho)d\Omega , 
\end{equation}
where
$\frac{d\sigma}{d\Omega} $                                    
contains the usual sum of amplitudes squared and phase space factors.
An alternative way would be to include Pauli blocking in the definition of
phase parameters \cite{ARS}.

\subsection{Results for $pp$ and $nn$ cross sections} 

As a baseline, we start by showing the $pp$ cross section as a function of the 
 momentum $q_0$ and at different densities of symmetric matter, see Fig.~2.
 Until otherwise noted, we use in-vacuum kinematics to define the total 2N         
 momentum in the nuclear matter rest frame (that is, the target nucleon is, on
 the average, at rest). 
 The range of momentum in Fig.~2 corresponds to values of the in-vacuum laboratory
 energy between approximately 20 and 260 MeV.                    
\begin{figure}
\begin{center}
\vspace*{0.3cm}
\hspace*{-0.5cm}
\psfig{figure=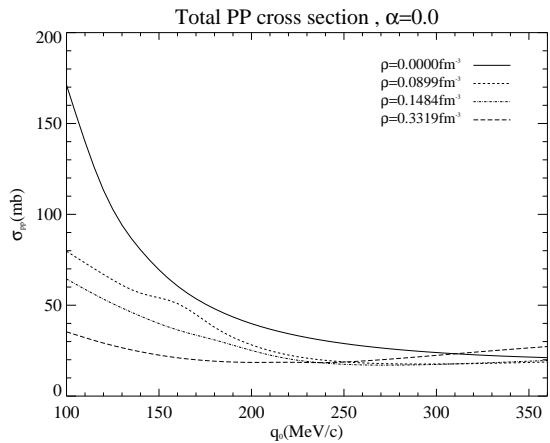,height=6.0cm}
\vspace*{0.3cm}
\caption{ Total $pp$ cross section in symmetric matter as a function of the 
 momentum in the 2N c.m. frame at the densities 
indicated in the figure. 
} 
\label{two}
\end{center}
\end{figure}

As compared to the predictions of Ref.~\cite{LM}, 
where the use of the $K$-matrix approach imposes unitarity of
the in-medium scattering matrix, 
the suppression of 
the cross section with increasing 
density is less pronounced. This was                                              
already pointed out in Ref.~\cite{Fuchs} 
and is most likely due to the different handling
of the unitarity issue as noted above.  
Overall, we find reasonable qualitative agreement with previous predictions.
We also notice that the high-density cross section tends to rise again at     
large momenta, a behavior already observed in Refs.~\cite{LM,Fuchs} particularly
for the $pp$ channel. 
                    
A simpler approach to calculating in-medium cross sections consists of scaling the  
free-space value according to the relation \cite{Gale}                               
\begin{equation}
\sigma(q_0,\rho) = (m^*(\rho)/m)^2 \sigma _{free}(q_0)
\end{equation}
where $m^*$ is the effective mass. 
In Fig.~3 we compare this approximation with the predictions from our full calculation. We see that 
at low momentum, and particularly at the lower densities, the agreement is quite
good, but it strongly deteriorates at higher momenta and densities. Medium effects on the 
interaction are clearly important, particularly the interplay between
the (energy-dependent) Pauli blocking mechanism and the dressing of the quasiparticle 
due to the surrounding medium.                                                

We next examine the 
role of isospin asymmetry, the focal point of this paper.                        
In Fig.~4, the ratio of the in-medium cross section to the free-space one
is shown for both $pp$ and $nn$ scatterings as a function of the asymmetry.
The total density is fixed to its value near saturation and the momentum $q_0$
is chosen to be equal to the corresponding Fermi momentum ($k_F$=1.3 $fm^{-1}$).
The dependence on $\alpha$ is generally weak, which may be attributed to the 
``competing'' roles of the effective mass and Pauli blocking. For the $pp$ case,
for instance, 
the smaller effective mass \cite{SBK} and the lower Fermi momentum (see Fig.~1) would tend to 
decrease and increase the cross section, respectively. The opposite happens
in the $nn$ case. For 
   the momentum considered in Fig.~4, the role of the effective mass appears to be
  dominant, lowering the $pp$ cross section and raising the $nn$ one.
  This trend of the $pp$/$nn$ cross sections versus $\alpha$ and relative to each
  other 
  is in qualitative agreement with predictions
  based on scaling the cross section as in Eq.(4), with effective masses 
  obtained from the  modified Gogny interation recently used in isospin-dependent
  BUU calculations \cite{Li05}.

 Figure~5 displays the cross section ratios for fixed values of the asymmetry
 and density but changing relative momentum. There, we see that the ratio approachs one  
 in the high-momentum limit.                     

 In Fig.~6, the cross section ratios are shown as functions of the       
 (average) Fermi momentum but fixed asymmetry and $q_0$. The latter is chosen
 as in Fig.~4. The cross sections are seen to go down rather quickly with
 increasing density but then rise again at the higher densities.                
  As we mentioned earlier, this was already observed,             
  particularly in the $pp$ channel, in 
 previous microscopic calculations of cross sections in 
 symmetric matter \cite{LM,Fuchs}. In this region of the phase space, 
   disagreement with predictions based on Eq.~(4) \cite{Li05} reflects
 the discrepancy already noted in the high-density part of the lower panel in Fig.~3.

\begin{figure}
\begin{center}
\vspace*{0.3cm}
\hspace*{-0.5cm}
\psfig{figure=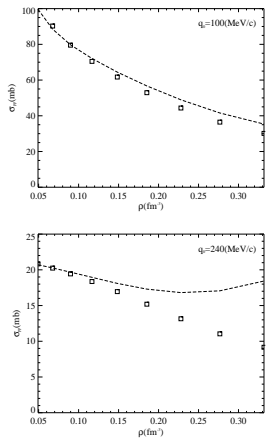,height=6.0cm}
\vspace*{0.3cm}
\caption{Our calculated total $pp$ cross section in symmetric matter                    
compared with the one obtained from Eq.~(4) (squares).
} 
\label{three}
\end{center}
\end{figure}

\begin{figure}
\begin{center}
\vspace*{0.3cm}
\hspace*{-0.5cm}
\psfig{figure=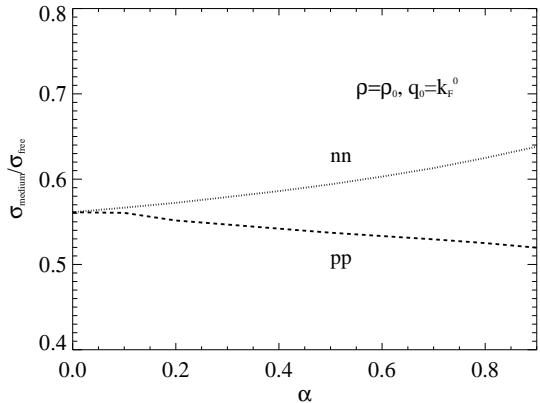,height=6.0cm}
\vspace*{0.3cm}
\caption{ Ratio of the total $pp$ and $nn$ cross sections to their free-space
values as a function of the asymmetry
near saturation density and fixed 2N relative momentum. 
} 
\label{four}
\end{center}
\end{figure}

\begin{figure}
\begin{center}
\vspace*{0.3cm}
\hspace*{-0.5cm}
\psfig{figure=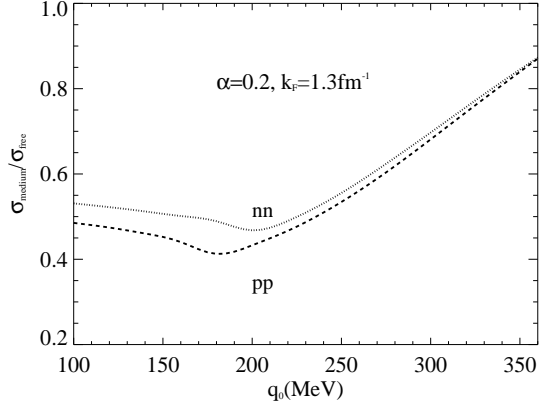,height=6.0cm}
\vspace*{0.3cm}
\caption{ Ratio of the total $pp$ and $nn$ cross sections to their free-space
values versus the momentum $q_0$ at fixed asymmetry and total density.          
} 
\label{five}
\end{center}
\end{figure}

\begin{figure}
\begin{center}
\vspace*{0.3cm}
\hspace*{-0.5cm}
\psfig{figure=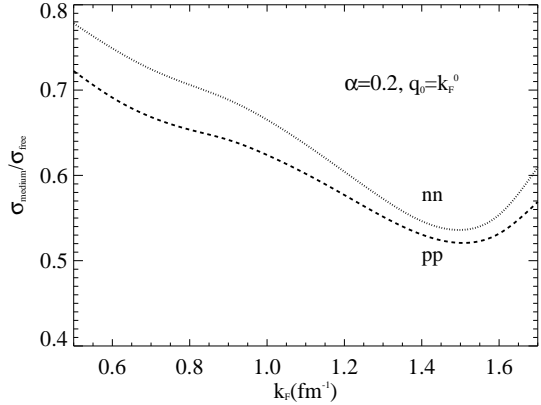,height=6.0cm}
\vspace*{0.3cm}
\caption{Ratio of the total $pp$ and $nn$ cross sections to their free-space
values versus the average Fermi momentum for fixed asymmetry and relative 
momentum.
} 
\label{six}
\end{center}
\end{figure}

Following up on the comments we made at the end of Subsection IIA, 
in the next figures we examine the impact of treating the 2N total momentum
with respect to nuclear matter, $P$, as an independent variable (rather than 
linked to $q_0$ through the simplifying assumption of in-vacuum kinematics).

In Fig.~7, we show $pp$ and
$nn$ cross section ratios under the same conditions of density and two-nucleon
relative momentum as in Fig.~4 but changing value of the total 2N                
momentum, $P$. The general tendency seems rather independent of $P$, although some 
minor differences can be seen with Fig.~4 as well as among the three situations 
displayed 
in Fig.~7. In particular, we notice that for the lower values of $P$              
the $pp$ cross section shows a lesser
tendency to decrease with increasing $\alpha$, and perhaps even a slight tendency
to rise again.                     
This may be understood in terms
of Pauli blocking, which becomes more important at lower values of $P$ and whose effect is opposite the one coming from 
 the decreasing proton mass.

\begin{figure}
\begin{center}
\vspace*{0.3cm}
\hspace*{-0.5cm}
\psfig{figure=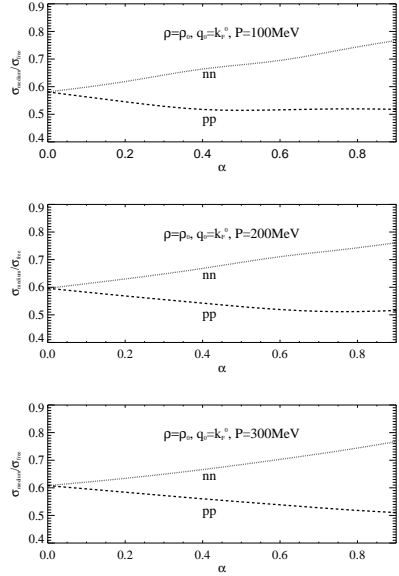,height=8.0cm}
\vspace*{0.3cm}
\caption{ Ratio of the total $pp$ and $nn$ cross sections to their free-space
values as a function of the asymmetry
at saturation density and fixed relative momentum. The value of the total 
momentum in the nuclear matter rest frame is indicated in each frame.
} 
\label{seven}
\end{center}
\end{figure}
      
Next, to broaden our previous analysis, we show in Fig.~8 and 
Fig.~9 situations analogous to those presented in 
 Fig.~4 and Fig.~7, respectively, but at a lower density.
The chosen value of the average Fermi momentum, which is also taken as the 
value of $q_0$, 
corresponds to approximately
one-half of saturation density.
Sensitivity to the asymmetry appears stronger at lower density, as seen
from the more pronounced differences between the $pp$ and $nn$ curves in both
Fig.~8 and Fig.~9.

\begin{figure}
\begin{center}
\vspace*{0.3cm}
\hspace*{-0.5cm}
\psfig{figure=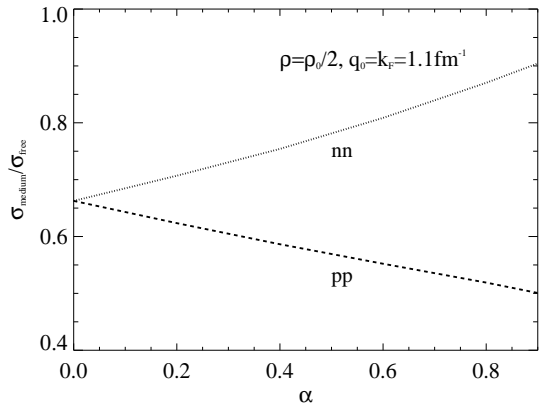,height=6.0cm}
\vspace*{0.3cm}
\caption{ Ratio of the total $pp$ and $nn$ cross sections to their free-space
values as a function of the asymmetry at approximately one half of 
 saturation density and fixed relative momentum. 
} 
\label{eight}
\end{center}
\end{figure}

\begin{figure}
\begin{center}
\vspace*{0.3cm}
\hspace*{-0.5cm}
\psfig{figure=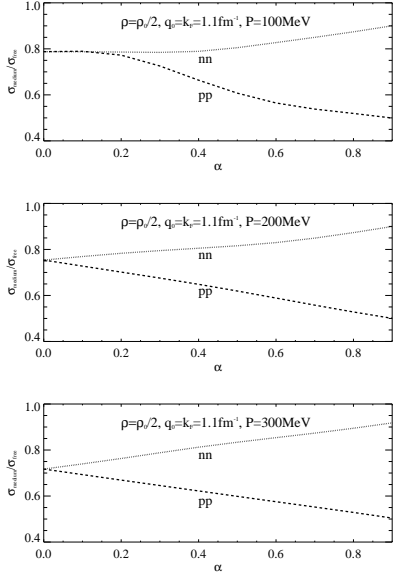,height=8.0cm}
\vspace*{0.3cm}
\caption{ Ratio of the total $pp$ and $nn$ cross sections to their free-space
values as a function of the asymmetry at approximately one-half of
 saturation density and fixed relative momentum. The value of the total 
momentum in the nuclear matter rest frame is indicated in each frame.
} 
\label{nine}
\end{center}
\end{figure}

\subsection{Results for $np$ cross sections} 

Similarly to what was done in Fig.~2 for the $pp$ case, 
we begin with showing in Fig.~10 the basic density/momentum dependence of the $np$ cross 
section in symmetric matter.                                          
(In-vacuum kinematics is used at this point.)
We notice here that the cross section at the density equivalent to 
$k_F$=1.1 $fm^{-1}$ shows 
some local enhancement which is considerably more 
pronounced than what can be seen in the $pp$ case,         
see Fig.~2. This comparison suggests that enhanced attraction is                 
generated at low density in the T=0 partial waves, most likely those with the 
quantum numbers of deuteron.

\begin{figure}
\begin{center}
\vspace*{0.3cm}
\hspace*{-0.5cm}
\psfig{figure=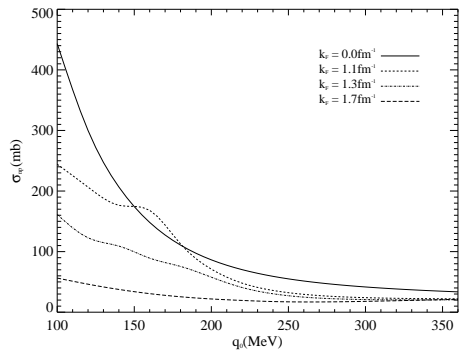,height=5.0cm}
\vspace*{0.3cm}
\caption{ Total $np$ cross section in symmetric matter as a function of the 
 momentum in the 2N c.m. frame at the densities 
indicated in the figure. 
} 
\label{ten}
\end{center}
\end{figure}

\begin{figure}
\begin{center}
\vspace*{0.3cm}
\hspace*{-0.5cm}
\psfig{figure=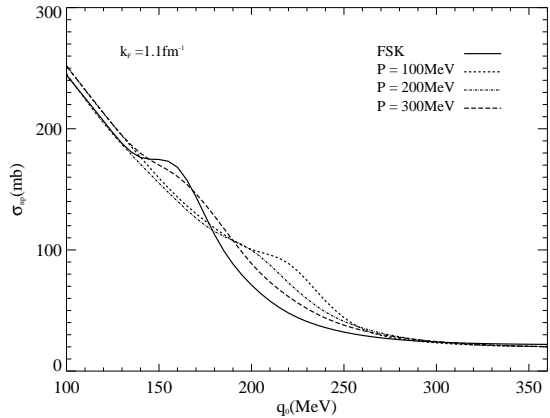,height=6.0cm}
\vspace*{0.3cm}
\caption{ Total $np$ cross section in symmetric matter at approximately one-half 
of saturation density under different kinematical conditions. See text for details.
} 
\label{eleven}
\end{center}
\end{figure}
In Fig.~11, we explore more closely the dependence of these ``structures" on the 
kinematics. The label ``FSK'' stands for free-space kinematics (as was used in Fig.~10),
whereas the other three curves refer to different choices of the total 
momentum $P$ as indicated in the figure.
Clearly, the location of these enhancements is determined by the kinematics,      
moving to lower values of $q_0$ for higher values of $P$. Although noticible, 
these gentle fluctuations do not resemble at all the extremely large resonance-like 
peaks reported in some earlier works \cite{ARS} and                          
interpreted as precursor of a superfluid phase transition in nuclear matter
at particular densities/momenta. Although crucial for generating those 
resonances in 
 Ref.~\cite{ARS} was the inclusion of hole-hole scattering, 
other authors have reported large bound-state signatures even without 
hole-hole contributions \cite{hh}. 
Our predictions do not confirm those 
findings. Very mild fluctuations are predicted in Ref.~\cite{Fuchs}, whereas
no enhancements at all appear in the calculations of      
Ref.~\cite{LM}.

We will focus next on the asymmetry dependence of the $np$ cross section.
Having observed 
in the previous subsection that asymmetry effects on protons and neutrons tend to
move the cross section in opposite directions, 
we may reasonably expect very mild dependence on $\alpha$. 
This is indeed confirmed in Fig.~12 (where in-vacuum kinematics is used), and 
Fig.~13 (where, as before, the sensitivity to the choice of $P$ is explored). 

For in-medium $np$ scattering, the nature of the cross section is the result of a  
subtle combination of multiple effects. These come from both the neutron and proton 
effective masses 
and the role of  both the neutron and proton Fermi                        
 momenta in the expression 
of the (asymmetric) Pauli operator \cite{AS02}. 
In the end, the behavior versus $\alpha$ is nearly flat, although a very slight tendency to rise can 
be identified, 
with the exception of the lower-$P$ curve in Fig.~13.

\begin{figure}
\begin{center}
\vspace*{0.3cm}
\hspace*{-0.5cm}
\psfig{figure=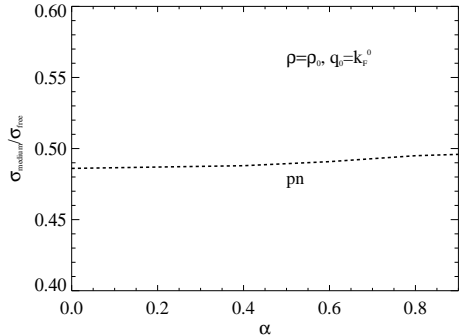,height=5.0cm}
\vspace*{0.3cm}
\caption{ Ratio of the total $np$ cross sections to its free-space
value as a function of the asymmetry
at approximately saturation density and fixed relative momentum. 
} 
\label{twelve}
\end{center}
\end{figure}
\begin{figure}
\begin{center}
\vspace*{0.3cm}
\hspace*{-0.5cm}
\psfig{figure=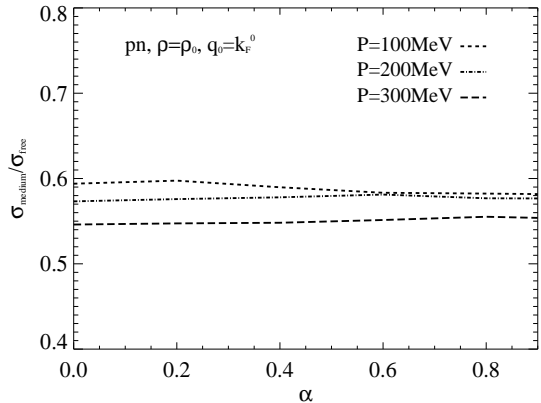,height=6.0cm}
\vspace*{0.3cm}
\caption{ Ratio of the total $np$ cross section to its free-space
value as a function of the asymmetry
at saturation density and fixed relative momentum. The value of the total 
momentum in the nuclear matter rest frame is indicated in each frame.
} 
\label{thirteen}
\end{center}
\end{figure}

\section{Conclusions} 
We have presented microscopic calculations of cross sections for scattering of 
nucleons in neutron-rich matter. 
The sensitivity to the asymmetry in neutron/proton ratio comes through    
the                                
combined effect of Pauli blocking and changing effective masses. 
The lowering(rising) of the proton(neutron) Fermi momentum and the reduced(increased)
proton(neutron) effective mass tend to move the cross section in opposite 
directions. 
In the phase space regions we have examined here, the general tendency is for the 
$pp$ cross section to go down and the $nn$ one to rise with increasing asymmetry.
This indicates that the role of the effective mass is dominant over
Pauli blocking. 

When both kinds of nucleons are involved, though, dispersive and 
Pauli blocking effects on each type of nucleon result in large cancelations.
The $np$ cross section is essentially 
$\alpha$-independent, although minor variations can be seen depending on the
kinematics. 

On the other hand, 
the basic density/momentum dependence of the $np$ cross section in symmetric matter
appears rather interesting. Different predictions disagree concerning the 
presence of large bound-state effects which might be associated with the onset 
of superfluidity.

In summary,          
sensitivity to the asymmetry can be non-negligible for scattering
of identical nucleons. The degree of sensitivity depends on the 
region of the energy-density-asymmetry phase space under consideration.
We conclude that 
 the proton mean free path could be affected in a significant way by in-medium 
 scattering, mostly through $\sigma _{pp}$. 
          
To test these findings, 
it is very useful to identify HI collision observables specifically 
sensitive to the two-body cross sections. Possible probes of the isospin-dependence
of the in-medium NN cross sections are presently being investigated \cite{MSU}.
Whereas traditional means such as measurements of the stopping power are found
to be ambiguous with respect to determining isospin dependence, potential 
candidates include isospin tracers like the neutron/proton free-nucleon ratio
in reactions induced by radioactive beams in inverse kinematics \cite{MSU}. 
\\ \\ 
\begin{center}
{\bf ACKNOWLEDGMENTS}
\end{center}
The authors        
acknowledge
financial support from the U.S. Department of Energy under grant number DE-FG02-03ER41270.


\begin{references}
\bibitem{BUU} See, for instance, B.A. Li, Phys. Rev. Lett. {\bf 85}, 4221 (2000), and references therein.
\bibitem{SBK} F. Sammarruca, W. Barredo, and P. Krastev, Phys. Rev. C {\bf 71},
064306 (2005). 
\bibitem{PP} V.R. Pandharipande and S.C. Pieper, Phys. Rev. C {\bf 45}, 791 (1992).
\bibitem{Gale} D. Persram and C. Gale, Phys. Rev. C {\bf 65}, 064611 (2002). 
\bibitem{LM} G.Q. Li and R. Machleidt, Phys. Rev. C {\bf 48}, 1702 (1993);
 {\bf 49}, 566 (1994). 
\bibitem{Fuchs} C. Fuchs, A. Faessler, and M. El-Shabshiry, Phys. Rev. C {\bf 64}, 024003 (2001).         
\bibitem{Glauber} See, for instance, R. Crespo and R.C. Johnson, Phys. Rev. C {\bf 60},
034007, (1999), and references therein. 
\bibitem{AS02} D. Alonso and F. Sammarruca, Phys. Rev. C {\bf 67}, 054301 (2003).         
\bibitem{Mac89} R. Machleidt, Adv. Nucl. Phys. {\bf 19}, 189 (1989). 
\bibitem{ARS} T. Alm, G. Ropke, and M. Schmidt, Phys. Rev. C {\bf 50}, 31 (1994). 
\bibitem{Li05} Bao-An Li, private communications.                    
\bibitem{hh} A. Bohnet, N. Ohtsuka, J. Aichelin, R. Linden, and A. Faessler, 
Nucl. Phys. {\bf A494}, 349 (1989). 
\bibitem{MSU} B.-A. Li, P. Danielewicz, and W. Lynch, nucl-th/0503038. 


\end{references}
\end{document}